\documentclass[doublecol]{epl2}

\title{RVB  signatures in the spin dynamics of the square-lattice Heisenberg antiferromagnet} 
\author{E. A. Ghioldi, M. G. Gonzalez, L. O. Manuel and A. E. Trumper}
\shortauthor{E. A. Ghioldi \etal}

\institute{Instituto de F\'{\i}sica Rosario (CONICET) and Universidad Nacional de Rosario, Boulevard 27 de Febrero 210 bis, 
(2000) Rosario, Argentina}

\abstract{
We investigate the spin dynamics of the square-lattice spin-$\frac{1}{2}$ Heisenberg antiferromagnet by means of an improved mean field Schwinger
boson calculation. By identifying both, the long range N\'eel and the RVB-like  components
of the ground state, we propose an educated guess for the mean field {magnetic} excitation consisting on a linear 
combination of local and bond spin flips to compute the dynamical structure factor. Our main result is that when this {magnetic} excitation 
is optimized in such a way that the corresponding sum rule is fulfilled, we recover the low and high energy spectral weight features of the experimental spectrum.
In particular, the anomalous spectral weight depletion at $(\pi,0)$ found in recent inelastic neutron scattering experiments can be attributed to the interference of the triplet bond excitations of the 
RVB component of the ground state. We conclude that the Schwinger boson theory seems to be a good candidate to adequately interpret the dynamic properties of the 
square-lattice Heisenberg antiferromagnet.}

\begin{document}

\maketitle

\section{Introduction}
The nature of the spin excitations in two dimensional (2D) quantum antiferromagnets (AF) represents
one of the major challenges in strongly correlated electron systems. 
Originally motivated by the cuprates superconductors \cite{Anderson87}, the square-lattice Heisenberg antiferromagnet has long been the prototypical model
to investigate the validity of  the spin wave \cite{Manousakis91} and the resonant valence bond (RVB)
descriptions \cite{Anderson01}. Here, in contrast to the one dimensional case \cite{Fadeev81,Nagler91,Tennant93,Mourigal13}, the ground state has long range N\'eel order, so it is expected that spin-$1$ magnon excitations take correctly 
into account the low energy part of the spectrum \cite{Manousakis91}. However, due to the presence of strong quantum fluctuations, there is an increasing belief that the high energy 
part of the spectrum can be described by pairs of spin-$\frac{1}{2}$ spinons, representing the excitations of the isotropic component of the ground state \cite{Anderson01}.\newline 
\indent Recent high resolution inelastic neutron scattering experiments performed  in the metal organic compound $Cu(DCOO)_2 \cdot 4D_2O$ (CFDT) --a known realization of the square-lattice Heisenberg AF model-- 
seem to support this argument \cite{DallaPiazza15}. Specifically, the spectrum has {\it i}) well defined low energy magnon peaks 
around $(\pi,\pi)$; {\it ii}) a wipe out of the intensity along with a 
downward renormalization of the dispersion near $(\pi,0)$ and {\it iii}) the continuum excitation at $(\pi,0)$ is isotropic, in contrast 
to the $(\frac{\pi}{2}, \frac{\pi}{2})$ one. Signals of these features were also found in the undoped cuprates, being their origin attributed to the 
possible presence of extra ring exchange interactions due to charge fluctuations effects \cite{Headings10}. In CFDT, however, the electrons are much more 
localized implying a negligible role of charge fluctuations {\cite{Christensen07}}. In fact, the  dispersion relation measured in CFDT has a deviation with respect to linear spin wave theory at 
$(\pi,0)$ but agrees very well with series expansion \cite{Zheng05} and quantum Monte Carlo \cite{Syljuasen00} calculations performed in the pure Heisenberg model. 
Furthermore, spin wave calculations reproduce the correct spectral weight of the spectrum around $(\pi,\pi)$ while the anomaly
at $(\pi,0)$ seems to be reproduced by a Gutzwiller projected wave function calculation 
where the isotropic continuum is interpreted as spatially extended pairs of fermionic spinons \cite{DallaPiazza15}. 
This calculation, however, fails to consistently describe at the same time the low energy part of the spectrum near $(\pi,\pi)$. 

On the other hand, recently, a whole description of the  dispersion relation in terms of magnons was performed using a continuous 
similarity transformation \cite{Powolski15}. {In this context, the downward renormalization at $(\pi,0)$ was attributed to 
a significant spectral weight transfer from the single magnon states to the three magnon continuum.} 
Unfortunately, the corresponding dynamical structure factor has not been computed; so a close comparison with 
the measured spectral weight has not been carried out yet.      

Alternatively, two decades ago, Arovas and Auerbach developed a bosonic spinon-based theory \cite{Arovas88}. Using a mean field Schwinger boson (MFSB) theory  they computed 
the dynamical structure factor $S({\bf k},\omega)$ 
in the square-lattice antiferromagnet (see fig. \ref{fig1}(a)). Even if the spectrum shows a low energy dominant spectral weight around $(\pi,\pi)$
with a magnon dispersion that matches the spin wave result, the spectral weight at $(\pi,0)$ and $(\frac{\pi}{2},\frac{\pi}{2})$ are practically the same, 
namely, at odds with the experimental spectrum (see fig. \ref{fig1}(b)). In view of the recent neutron scattering experiments on CFDT, and the theoretical 
difficulties mentioned above, the search for a consistent theoretical description that takes into account 
the main features of the spectrum has become an important goal.

In this paper. we perform an improved mean field Schwinger boson calculation of the dynamical structure factor for the square-lattice Heisenberg antiferromagnet.
Using the fact that the mean field ground state can be described by a N\'eel and an {\it averaged} RVB component, we investigate the corresponding spectral properties and
 propose an educated guess for the {magnetic} excitations  
consisting of a linear combination of local and bond spin flips. Notably, when this {magnetic} excitation is optimized in such a way that the sum rule 
$\int d\omega \sum_{\bf k}S({\bf k},\omega)= NS (S+1)$ is fulfilled
we find that the main spectral weight features of the experimental spectrum are reproduced quite well (see fig. \ref{fig3}(a)).
In particular, our results support the idea that the anomalous spectral weight depletion at $(\pi,0)$ can be attributed to the interference of the 
triplet bond excitations corresponding to the {\it averaged} RVB component of the ground state \cite{Christensen07}.

\section{N\'eel and RVB components of the mean field Schwinger boson ground state}

It is firmly established that the ground state of the spin-$\frac{1}{2}$ Heisenberg model on the square-lattice 
is an $SU(2)$ broken symmetry quantum N\'eel state \cite{Manousakis91}. Nonetheless, the nature of the zero point quantum fluctuations and the spin 
excitations above the ground state are still controversial. 
It has been proposed that the zero point quantum fluctuations have 
both local and RVB character, the latter being related to the anomaly found at $(\pi,0)$ in the neutron scattering experiments of CFDT \cite{Christensen07}. 
In this section we show that the ground state provided by the MFSB can be related to the above proposal. \\

 Here, we present the main steps of the mean field Schwinger boson theory, 
 originally developed by Arovas and Auerbach\cite{Arovas88}.
Within the Schwinger boson representation the spin operators are expressed as
${\hat{\bf S}}_i\!\!=\! \frac{1}{2}{\bf b}^{\dagger}_i \vec{\sigma} \;{\bf b}_i$, with the spinor
${\bf b}^{\dagger}_i \!=\!(\hat{b}^{\dagger}_{i\uparrow}; \hat{b}^{\dagger}_{i\downarrow})$ composed by the bosonic 
operators $\hat{b}^{\dagger}_{i\uparrow}$ and $\hat{b}^{\dagger}_{i\downarrow}$, and
$\vec{\sigma}\!\!=\!\!(\sigma^x,\sigma^y,\sigma^z)$ the Pauli matrices. To fulfill the spin algebra the constraint 
of $2S$ bosons per site, $\sum_{\sigma}\hat{b}^{\dagger}_{i\sigma}\hat{b}_{i\sigma}\!=2S$, must be imposed. Using this representation
the AF Heisenberg Hamiltonian results \cite{Arovas88}, 

\begin{equation}
\frac{J}{2} \sum_{<i,j>} \hat{\bf S}_i \!\cdot\! \hat{\bf S}_j= \frac{J}{2} \sum_{<i,j>} \left [ S^2-2\hat{A}^{\dagger}_{ij}\hat{A}_{ij}\right ],
\label{int}
\end{equation}

\noindent where $J>0$ is the exchange interaction between nearest neighbors $<\!\!ij\!\!>$ and  
$\hat{A}^{\dagger}_{ij}\!=\!\frac{1}{2}\sum_{\sigma}\sigma \hat{b}^{\dagger}_{i \sigma}
\hat{b}^{\dagger}_{j \bar{\sigma}}$ is a singlet bond operator \cite{note}.
Introducing a Lagrange multiplier $\lambda $ to impose the 
local constraint on average  and performing a mean field decoupling of 
eq. (\ref{int}), such that $A_{ij}=\langle\hat{A}_{ij}\rangle=\langle\hat{A}^{\dagger}_{ij}\rangle$, the diagonalized mean field Hamiltonian 
yields\cite{Arovas88,Ceccatto93,Mezio11}
$$\hat{H}_{MF}= E_{\texttt{gs}}+\sum_{\bf k} \omega_{\bf k} \left[\hat{ \alpha}^{\dagger}_{{\bf k}\uparrow} 
\hat{\alpha}_{{\bf k}\uparrow}+
\hat{\alpha}^{\dagger}_{-{\bf k}\downarrow} \hat{\alpha}_{-{\bf k}\downarrow} \right],$$
where 
$$E_{\texttt{gs}}=\sum_{\bf k}\omega_{\bf k}{- \frac{N}{2} \sum_{\delta} J(S^2-A^2_{\delta})}-\lambda N (2S+1)$$
is the ground state energy, {with $A_{\delta}\!$ chosen real 
and $\delta$ connecting all the first neighbors of a square lattice, while}   
$$
\omega_{{\bf k}\uparrow}=\omega_{{\bf k}\downarrow}= \omega_{\bf k}=[\lambda^2- 
(\gamma^A_{\bf k})^2]^{\frac{1}{2}},
$$
is the spinon dispersion relation with  
$\gamma^A_{\bf k}\!\!=\!\! \frac{1}{2} J\sum_{\delta} A_{\delta} \sin ({\bf k}. \delta)$,  
The ground state wave function $|\texttt{gs} \rangle$ is such that 
$\hat{ \alpha}_{{\bf k}\uparrow} |\texttt{gs}\rangle =\hat{ \alpha}_{{\bf k}\downarrow} |\texttt{gs}\rangle=0$ and has a Jastrow form \cite{Chandra90}

\begin{equation}
|\texttt{gs}\rangle\; = e^{\sum_{\bf k} f_{\bf k} b^{\dagger}_{{\bf k} \uparrow} b^{\dagger}_{-{\bf k} \downarrow}}
|0\rangle_b,
\label{WF}
\end{equation}

\noindent where $|0\rangle_b$ is the vacuum of Schwinger bosons $b$ and  $f_{\bf k}= -v_{\bf k}/u_{\bf k}$, with  $u_{\bf k}\!= \![\frac{1}{2}(1+
\frac{\lambda}{\omega_{\bf k}} )]^{\frac{1}{2}}$ and  $v_{\bf k}\!= \!\imath\ {\it sgn}(
\gamma^A_{\bf k})[\frac{1}{2}(-1+\frac{\lambda}{\omega_{\bf k}} )]^{\frac{1}{2}}$ the Bogoliubov coefficients used to diagonalize
 $\hat{H}_{MF}$. The self consistent mean field equations for $A_{\delta}\!$ and $\lambda$ are,

\begin{eqnarray}
A_{\delta}&=& \frac{1}{2N}\sum_{\bf k} \frac{\gamma^A_{\bf k}}{\omega_{\bf k}} \sin ({\bf k}. {\delta})  \label{self}\\
S+\frac{1}{2}&=& \frac{1}{2N}\sum_{\bf k} \frac{\lambda}{\omega_{\bf k}} \label{const}. 
\end{eqnarray}
The rotational invariance preserved by the MFSB allows us to study finite size systems directly, 
avoiding the singularities that arise when a broken symmetry state is assumed, like in spin wave theory \cite{Zhong93}. 
Numerical computation of the above self consistent equations shows two gauge related singlet solutions, the $s$ wave ($A_{\delta_x}=A_{\delta_y}$) \cite{Arovas88} and the $d$ wave 
($A_{\delta_x}=-A_{\delta_y}$) \cite{Yoshioka89} solutions. In particular, 
as soon as the system size $N$ increases, both solutions develop $180^{\circ}$ N\'eel 
correlations signaled by the minimum gap of the spinon dispersion at $\pm(\frac{\pi}{2},\frac{\pi}{2})$ that vanishes  in the thermodynamic limit. 
This closing of the gap is related to the spontaneous $SU(2)$ broken symmetry N\'eel state \cite{Sarker89,Chandra90}. By the way, it is instructive 
to rearrange eq. (\ref{const}) as 

\begin{equation}
 S=\frac{1}{N} \sum_{\bf k} \frac{|f_{\bf k}|^2}{[1-|f_{\bf k}|^2]}.
\label{meaning}
\end{equation}

\noindent In the thermodynamic limit the self consistent solutions yield $|f_{\pm(\frac{\pi } {2},\frac{\pi } {2})}|\rightarrow 1$. Then, the quantum corrected 
magnetization $m$ can be obtained as the singular part of eq. (\ref{meaning}) \cite{Chandra90}, 
\begin{equation}
m= \frac{2}{N} \frac{|f_{(\frac{\pi }{2},\frac{\pi}{2})}|^2}{[1-|f_{(\frac{\pi }{2},\frac{\pi}{2})}|^2]}.
\label{eme}
\end{equation}
\noindent From eq. (\ref{WF}), it is clear that the singular behavior is due to  the condensation of spin up and spin down bosons at 
${\bf k}=\pm(\frac{\pi }{2},\frac{\pi}{2})$. Consequently, {in the thermodynamic limit,} the ground state can be splitted as

$$
 |\texttt{gs}\rangle\;=|\texttt{c}\rangle |\texttt{n}\rangle,     
$$

\noindent where 

\begin{equation}
 |\texttt{c}\rangle= e^{{{\sqrt{\frac{N m}{2}}}} \left({b}^{\dagger}_{(\frac{\pi } {2},\frac{\pi } {2}) \uparrow} + {b}^{\dagger}_{-(\frac{\pi } {2},\frac{\pi } {2}) \uparrow} + 
i {b}^{\dagger}_{(\frac{\pi } {2},\frac{\pi } {2}) \downarrow} -i{b}^{\dagger}_{-(\frac{\pi } {2},\frac{\pi } {2}) \downarrow}\right)}|0\rangle_b\\
\end{equation}

\noindent is the condensate part which represents the quantum corrected N\'eel state, $\langle \texttt{C}|S^z_{\bf r}  |\texttt{C}\rangle= (-1)^{r_x+r_y} m $,   and 

$$
 |\texttt{n}\rangle\;=e^{ \small{\sum_{\small{{\bf k}\neq\pm (\frac{\pi } {2},\frac{\pi } {2})}} f_{\bf k} 
b^{\dagger}_{{\bf k} \uparrow} b^{\dagger}_{-{\bf k} \downarrow}} }
|0\rangle_b
$$

\noindent is the isotropic normal fluid part which represents the zero point quantum fluctuations of the ground state \cite{Chandra90}. 
A practical advantage of the MFSB theory is that, even working on finite systems, it is possible to keep track of the 
putative magnetic order. For instance, a finite size scaling of eq. (\ref{eme}) gives $m=0.3034$, 
in agreement with Arovas and Auerbach result \cite{Arovas88}.
Going back to real space the normal fluid part is, approximately,   

\begin{equation}
|\texttt{n}\rangle\; \approx e^{\sum_{ij} f_{ij} \hat{A}^{\dagger}_{ij} }
|0\rangle_b,
\label{WFreal}
\end{equation}

\noindent where $f_{ij}$ is the Fourier transform of $f_{\bf k}$. Equation (\ref{WFreal}) shows explicitly the singlet bond structure 
of the normal fluid part of the ground state. Although it is not a true RVB state, because the constraint is only satisfied on 
average, one can still interpret the normal fluid component of the ground state 
as an {\it averaged} RVB component.\\

Keeping in mind this picture for the ground state,   
two kind of {magnetic} excitations can be envisaged within the MFSB: 
local spin flips and bond spin flips acting on the condensate and normal fluid components, respectively. On one hand, magnonic-like excitations are created by 
$\hat{S}{^z}_{\bf q}$, which is a linear 
combination of the local operator $\hat{S}{^z}_{i}$. On the other hand, triplet bond excitations of the {\it averaged} 
RVB component can be created by  $D^{\dagger}_{\bf q}$, which is a linear 
combination of the bond operator

$$\hat{D}^{\dagger}_{i}=\frac{1}{2} (D^{\dagger}_{i\delta_x}+D^{\dagger}_{i\delta_y}), $$
 with
$$D^{\dagger}_{i\delta}= b^{\dagger}_{i\uparrow}b^{\dagger}_{i+\delta \downarrow}+b^{\dagger}_{i\downarrow}b^{\dagger}_{i+\delta \uparrow}$$

\noindent {that creates a triplet of $z-$component equal to zero;
while  operators like} 
$T^{\dagger}_{i\delta \uparrow}= b^{\dagger}_{i\uparrow}b^{\dagger}_{i+\delta \uparrow}$ and 
$T^{\dagger}_{i\delta \downarrow}= b^{\dagger}_{i\downarrow}b^{\dagger}_{i+\delta \downarrow}$
{create triplets of $z-$component equal to $\pm 1$, respectively.}
  Actually, other kind of triplet bond excitations such as $C^{\dagger}_{i\delta}= b^{\dagger}_{i\uparrow}b_{i+\delta
 \uparrow}-b^{\dagger}_{i\downarrow}b_{i+\delta \downarrow}$, 
 can be constructed. However, after a systematic study of all possible triplet bond excitations we have found that the correct spectral 
properties are recovered once the excitations {corresponding to $D^{\dagger}_{i\delta}$, $T^{\dagger}_{i\delta \uparrow}$, and  
$T^{\dagger}_{i\delta \downarrow}$} are incorporated in the dynamical structure factor calculation (see below).\\ 

\section{Dynamical structure factor study}
 
In this section we show the difficulty of the original MFSB theory \cite{Arovas88} to recover the anomaly of the spectrum at $(\pi,0)$  
and we propose an improved calculation  of the dynamical structure factor by considering, explicitly, 
the triplet bond excitations $\hat{D}^{\dagger}_{\bf q}$ mentioned in the previous section. At zero temperature the dynamical structure factor is
defined as
\begin{equation}
S({\bf k},\omega)= \sum_{n}  |\langle\texttt{gs}|\hat{\bf S}_{\bf k}|n\rangle|^2 \delta 
(\omega-(\epsilon_n-E_{\texttt{gs}}))\nonumber, 
\label{Dsf}
\end{equation}
where $|n\rangle$ are the spin-$1$ excited states. Plugging the corresponding mean field states in eq. (\ref{Dsf}) results in 
\begin{equation}
S({\bf k},\omega)\!=\!\frac{1}{4N}\!\!\sum_{{\bf q}} |u_{{\bf k}+{\bf q}} v_{\bf q} - u_{{\bf q}} v_{{\bf k}+{\bf q}}|^2 
\delta (\omega-(\omega_{-{\bf q}}+\omega_{{\bf k}+{\bf q}})).
\label{Skw}
\end{equation}
\begin{figure}
\begin{center}
\includegraphics[width=8.0cm]{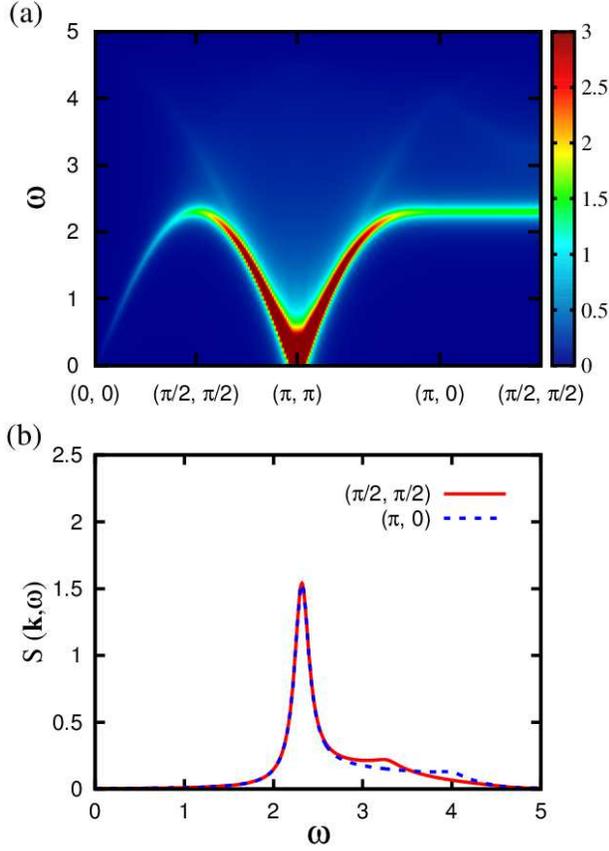}
\caption{Dynamical structure factor $S({\bf k},\omega)$ within the MFSB (eq. (\ref{Skw})). (a) intensity curve along a path of the Brillouin zone.
 (b) comparison between ${\bf k}=(\frac{\pi}{2},\frac{\pi}{2})$ (solid line) and ${\bf k}=(\pi,0)$ (dashed line).  }
\label{fig1}
\end{center}
\end{figure}

\begin{figure}
\begin{center}
\includegraphics[width=8.0cm]{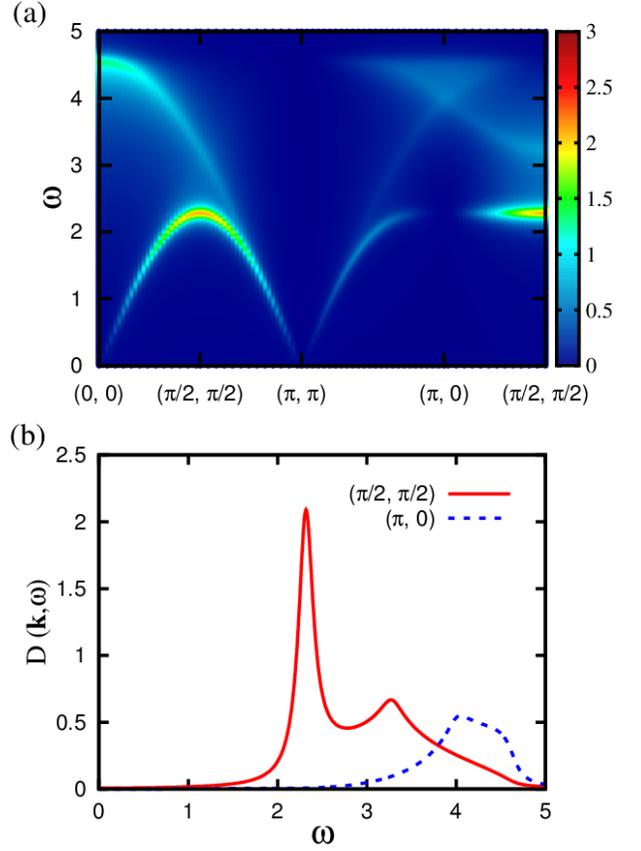}
\caption{Dynamical structure factor $D({\bf k},\omega)$ within the MFSB (eq. (\ref{dkw})). (a) intensity curve along a path of the Brillouin zone.
 (b) comparison between ${\bf k}=(\frac{\pi}{2},\frac{\pi}{2})$ (solid line) and ${\bf k}=(\pi,0)$ (dashed line).}
\label{fig2}
\end{center}
\end{figure}

\noindent By exploiting the fact that the MFSB theory fulfills the Mermin Wagner 
theorem\cite{Mermin66}, Arovas and Auerbach
 accessed to the above zero temperature dynamical structure factor coming from the finite temperature regime \cite{Arovas88}.
Here, instead, we work at zero temperature and use finite size systems to compute eq. (\ref{Skw})  
where the contribution of the $x,y$ and $z$ components are identical due to the $SU(2)$ symmetry of 
the ground state. However, due to the development of long range N\'eel order, the spectrum shows Bragg peaks and low energy 
Goldstone modes at ${\bf k}=(0,0)$ and $(\pi,\pi)$. This is shown  in fig. \ref{fig1}(a) where eq. (\ref{Skw}) is displayed in an intensity curve. 
 Even if the spectrum is expressed in terms of free pairs of spinons it is expected that gauge fluctuations, dynamically generated, will confine them 
into magnonic excitation \cite{Fradkin79, Starykh94}.
In fact, a simple first order calculation in perturbation theory supports the picture of tightly bond pair of spinons in the neighborhood of the Goldstone 
modes \cite{Mezio12}. 
Furthermore, the spectrum shows a low energy dominant spectral weight around $(\pi,\pi)$
with a magnon dispersion that matches the spin wave result. However the spectral weight at $(\pi,0)$ and $(\frac{\pi}{2},\frac{\pi}{2})$ are 
practically the same (see fig. \ref{fig1}(b)), 
that is, at odds with the anomaly observed at $(\pi,0)$ in the neutron scattering experiments of CFDT \cite{DallaPiazza15}. 
This anomaly was previously interpreted as a quantum mechanical interference 
due to the entanglement of the RVB component of the ground state \cite{Christensen07}. This motivated us to focus on the spectral properties of the triplet bond operator 
$D^{\dagger}_i$ which, within the context of our approximation, represents a proper magnetic excitation of the {\it averaged} RVB component
of the mean field ground state. The corresponding dynamical structure factor is 

\begin{equation}
D({\bf k},\omega)= \sum_{n}  |\langle\texttt{gs}|\hat{D}_{\bf k}|n\rangle|^2 \delta 
(\omega-(\epsilon_n-E_{\texttt{gs}}))\nonumber, 
\label{Dkw}
\end{equation}

\noindent with $\hat{ D}_{\bf k}$ the Fourier transform of $D^{\dagger}_i$. Within the MFSB eq. (\ref{Dkw}) results

\begin{equation}
D({\bf k}, \omega)\!=\!\frac{1}{N} \! \sum_{\bf q}[\left( \gamma_{\bf k} + \gamma_{\bf k + q}  \right)u_{\bf k} u_{\bf k + q}]^2  \delta\! \left( \omega - \omega_{\bf k} - \omega_{\bf k+q} \right)
\label{dkw}
\end{equation}

\noindent with  $\gamma_{\bf k}\!=\! \frac{1}{2} \sum_{\delta} \cos {\bf k}. \delta$.
 In fig. \ref{fig2}(a) is plotted eq. (\ref{dkw}) in an intensity curve. As expected, the spectral weight 
of these triplet bond fluctuations is mostly located at high energies. Notice the high energy spectral weight transfer with respect to $S({\bf k},\omega)$ 
(see fig. \ref{fig1}(a)). In particular, at $(\pi,0)$ the spectral weight transfer is complete (see fig. \ref{fig2}(b)). It is important to note that among all possible triplet bond excitations mentioned in the previous section 
 only the spectrum of $D({\bf q}, \omega)$ shows this interference at $(\pi,0)$.

In principle, such anomaly should appear in a rigorous calculation of $S({\bf k},\omega)$. But it is  known that,  
due to the relaxation of the local constraint, the excitations created by $\hat{S}_{\bf q}$ in the MFSB
are not completely physical, producing a factor $\frac{3}{2}$ in the sum rule, $\int d\omega \sum_{\bf k}S({\bf k},\omega)=\frac{3}{2} NS(S+1)$ \cite{Arovas88}.
Therefore, to improve the  $S({\bf k},\omega)$ calculation 
one should  project the mean field ground state and the spin-$1$ states onto the physical Hilbert 
space $\langle\texttt{gs}|P^{-1}{\bf S}_{\bf k}P|n\rangle$, an operation that is very difficult to implement  not only analytically \cite{Raykin93,Wang00} but 
numerically \cite{Chen93,Yoshioka95,Sorella12}. {In our case, the proper way to correct the mean field, or saddle point approximation, is to include 
Gaussian fluctuations of the mean field parameters {\cite{Trumper97}}; although its concrete computation for the dynamic structure factor may turn out a quite long task that is out of the scope of the present work \cite{Momoi13}.}
Alternatively, we look for an {\it effective} {magnetic} excitation that somehow mimics the {effect of the above projection}. To do that, {based on the nature of MFSB ground state discussed in the previous section,} we {propose} an educated guess for the {magnetic} 
excitation consisting on a linear combination $(1-\beta) {S}^z_{\bf k}+\beta D^{\dagger}_{\bf k}$ in such a way that the free parameter
$\beta$ can be adjusted to enforce the correct sum rule. 
Here it is important to note that both, $\hat{S}{^z}_{\bf q}$ and $\hat{D}^{\dagger}_{\bf q}$, produces the same change  
in the $z$-component of  the total spin, $\Delta S^{z}_{total}=0$, when they are applied to the ground state. 
 Then, the modified dynamical structure factor yields

\begin{equation}
S(\beta,{\bf k},\omega)\!=\! {3}\!\sum_{n} \! |\langle\texttt{gs}|(1-\beta) {S}{^z}_{\!\!\bf k}+\beta D^{\dagger}_{\bf k} |n\rangle|^2 \delta 
(\omega-(\epsilon_n-E_{\texttt{gs}}))\nonumber, 
\label{SD}
\end{equation}

\noindent {where the factor 3 is due to rotational invariance.} Notice that $\beta=0$ is equivalent 
to eq. (\ref{Skw}); while $\beta=1$ corresponds to eq. (\ref{dkw})	{, up to a factor $3$ }. After a little of algebra,  

\begin{eqnarray}
S(\beta\!,\!{\bf q}, \omega)\!\!\!\!\!&=&\!\!\!\!\!\frac{3\left(1- \beta\right)^2}{4N} \;  \sum_{\bf k} \Omega_{\;\bf k,\bf q}^2 \ \delta \left( \omega - \omega_{\bf k} - \omega_{\bf k+q} \right) + \nonumber\\
 &   +&\!\!\!\!\! \frac{{3}\beta^2}{N}   \; \sum_{\bf k} \Gamma_{\bf k,\bf q}^2  \; \delta \left( \omega - \omega_{\bf k} - \omega_{\bf k+q} \right) + \label{mixto}\\
 &  +&\!\!\!\!\! \frac{{3}\left(1- \beta\right) \beta}{N} \;  \sum_{\bf k} \Gamma_{\bf k,\bf q} \;  \Omega_{\;\bf k,\bf q} \ \delta \left( \omega - \omega_{\bf k} - \omega_{\bf k+q} \right) \nonumber
\end{eqnarray}
 
\noindent with
$$\Gamma_{{\bf k},{\bf q}}=|u_{{\bf k}+{\bf q}} v_{\bf q} - u_{{\bf q}} v_{{\bf k}+{\bf q}}|$$ and $$\Omega_{{\bf k},{\bf q}}=[\left( \gamma_{\bf k} + \gamma_{\bf k + q}  \right)u_{\bf k} u_{\bf k + q}].$$

\begin{figure}[t]
\begin{center}
\includegraphics[width=8.0cm]{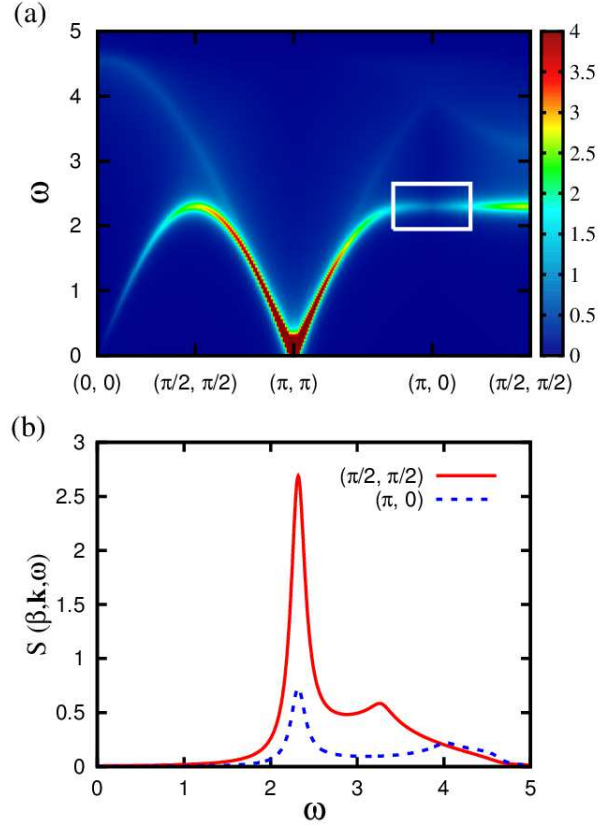}
\caption{Modified dynamical structure factor $S(\beta,{\bf k},\omega)$ within the MFSB (eq. (\ref{mixto})). For $\beta^*={0.315}$ the sum rule is fulfilled (eq. (\ref{sumrule})). (a) intensity curve along a path of the Brillouin zone. The white box shows the anomalous behavior at $(\pi,0)$ . (b) comparison between ${\bf k}=(\frac{\pi}{2},\frac{\pi}{2})$ (solid line) and ${\bf k}=(\pi,0)$ (dashed line). }
\label{fig3}
\end{center}
\end{figure}

\noindent We have computed eq. (\ref{mixto}) finding that the correct sum rule

\begin{equation}
\int\! \sum_{{\bf k}}S(\beta,{\bf k},\omega)d\omega= NS(S+1)
\label{sumrule}
\end{equation} 
 
 is fulfilled for
$\beta^{*}=0.315$. Notably,  $S(\beta^{*},{\bf q}, \omega)$ reproduces qualitatively quite well the low and the high energy features of the  expected spectrum. This is shown in fig. \ref{fig3}(a) where it is clear that the partial depletion of spectral weight at $(\pi,0)$ (white box of fig. \ref{fig3} (a)) is directly related to the presence of the triplet bond spin flips (see $S({\bf k},\omega)$ of fig. \ref{fig1}(a))). Therefore, we can conclude that the modified dynamical structure factor $S(\beta^{*},{\bf k},\omega)$ corresponding to the optimized operator $(1-\beta^{*}) {{S}^z}_{\bf k}+\beta^* D^{\dagger}_{\bf k}$ is mimicking the aimed effect of 
the projection operation; although, actually, we are not strictly imposing the local constraint. However, the fact that the expected features of the spectrum are recovered 
once the sum rule is fulfilled is encouraging. So far the dynamical structure factor results for $D({\bf k},\omega)$ and  $S(\beta^{*},{\bf k},\omega)$ 
correspond to the $s$ wave solution 
($A_{\delta_x}=A_{\delta_y}$) of the ground state. We have checked that if the $d$ wave solution ($A_{\delta_x}=-A_{\delta_y}$) is used 
the results are the same if the triplet bond operator is changed to $\hat{D}^{\dagger}_{i}=\frac{1}{2} 
(D^{\dagger}_{i\delta_x}-D^{\dagger}_{i\delta_y}) $. This shows the sensitivity of the MFSB to capture the intimate connection between the 
structure of the ground state and the corresponding triplet {bond} excitation. \\
Finally, it is important to point out that the anomaly at $(\pi,0)$ 
is only noticeable for the quantum spin case $S=1/2$. In fact, we have found that, as soon as $S$ is increased, the contribution of the bond spin flips to $S(\beta^{*},{\bf k},\omega)$ becomes negligible with respect to  the local spin flip one, thus recovering the expected large $S$ spin wave result.

\section{Summary and concluding remarks}

Motivated by the recent inelastic neutron scattering experiments performed in CFDT \cite{DallaPiazza15}, the experimental realization of the square-lattice quantum antiferromagnet, we have investigated the anomaly found in the spectrum at $(\pi,0)$ using mean field Schwinger bosons. Based on the proposal that such anomaly could be due to 
the quantum mechanical destructive interference of the RVB component of the ground state \cite{Christensen07}, we have studied the spectrum by exploiting the ability of the MFSB 
to properly describe the {magnetic} excitations above the N\'eel and the {\it averaged} RVB components of the ground state. In particular, we have  
found that the triplet $D^{\dagger}_{\bf k}$ bond spin excitations, the natural excitations above the {\it averaged} RVB, have an anomalous spectral property at
 $(\pi,0)$ that can be associated to the above mentioned interference. Then, in order to improve the original dynamical structure factor $S({\bf k},\omega)$ \cite{Arovas88} we have proposed a combined {magnetic} excitation $(1-\beta) {S}{^z}_{\bf k}+\beta D^{\dagger}_{\bf k}$ that gives rise to a modified dynamical structure factor $S(\beta,{\bf k},\omega)$. Remarkably, once it is optimized to enforce the sum rule, the main features of the spectrum at low and high energies are reproduced quite well. Unfortunately the optimized $S(\beta^{*},{\bf k},\omega)$ does not recover the rotonic feature at $(\pi, 0)$ but  this is because 
the local constraint is not imposed  exactly. If it is imposed carefully \cite{Chen93}, as it has also been done in the fermionic spinon case \cite{DallaPiazza15}, 
the rotonic features will be recovered. Regarding the possibility that free bosonic spinons can survive, or not, at $(\pi,0)$ is an issue that is beyond the scope of the present work. However, we think that the present results are calling for a more sophisticated calculation of the dynamical structure factor within the context of the Schwinger boson formalism, such as variational Monte Carlo \cite{Chen93, Yoshioka95, Sorella12} or 1/N correction \cite{Arovas88, Raykin93, Trumper97, Momoi13}. Work in the latter direction is in progress.

\acknowledgments
We thank A. Lobos for the careful reading of the manuscript and H. M. R{\o}nnow for useful discussions. This work was supported by  CONICET (PIP2012) under grant  Nro 1060.


\begin{thebibliography}{0}
\bibitem{note} For frustrated AF it has been discussed 
in the literature the necessity of using a second singlet bond operator\cite{Ceccatto93,Flint08,Mezio11} but for the unfrustrated case it is enough 
the original formulation with only the singlet bond operator 
$\hat{A}^{\dagger}_{ij}$.
\end{thebibliography}
\end{document}